\documentclass[aip,amsmath,amssymb,reprint,twocolumn]{revtex4-2}
\usepackage[T1]{fontenc}
\usepackage[latin9]{inputenc}
\usepackage{color}
\usepackage{amsmath}
\usepackage{amssymb}
\usepackage{graphicx}
\usepackage{color, colortbl}
\usepackage[pdftex,unicode=true,pdfusetitle,
 bookmarks=true,bookmarksnumbered=false,bookmarksopen=false,
 breaklinks=false,pdfborder={0 0 1},backref=false,colorlinks=false]{hyperref}
\usepackage[symbol]{footmisc}

\begin{document}
\title{Pressure induced electronic transitions in Samarium monochalcogenides}
\author{Debalina Banerjee}
\email[Corresponding author: ]{debalina.banerjee@kcl.ac.uk}
\affiliation{King's College London, The Strand, London, WC2R 2LS, UK}
\affiliation{National Physical Laboratory, Teddington, TW11 0LW, United Kingdom}
\author{Evgeny Plekhanov}
\affiliation{King's College London, The Strand, London, WC2R 2LS, UK}
\author{Ivan Rungger}
\affiliation{National Physical Laboratory, Teddington, TW11 0LW, United Kingdom}
\author{Cedric Weber}
\affiliation{King's College London, The Strand, London, WC2R 2LS, UK}
\date{\today}

\begin{abstract}
Pressure induced isostructural insulator to metal transition for SmS is characterised by the presence of an intermediate valence state at higher pressure which cannot be captured by the density functional theory. As a direct outcome of including the charge and spin fluctuations incorporated in dynamical mean field theory, we see the emergence of insulating and metallic phases with increasing pressure as a function of changing valence. This is accompanied by significantly improved predictions of the equilibrium lattice constants and bulk moduli for all Sm-monochalcogenides verifying experiments. Nudged Elastic Band analysis reveals the insulating states to have a finite quasiparticle weight, decreasing as the gap closes rendering the transition to be not Mott-like, and classifies these materials as correlated band insulators. The difference between the discontinuous and continuous natures of these transitions can be attributed to the closeness of the sharply resonant Sm-4f peaks to the fermi level in the predicted metallic states in SmS as compared to SmSe and SmTe.
\end{abstract}

\maketitle

An existing present day challenge is to invent a feasible successor to the CMOS technology oriented towards increasing computer clock speeds and power performances whose trends, as per Moore's law\cite{MooresLaw}, have saturated since 2003. This requires arresting of the switching power, steering us towards a different form of technology with nano-scalability where a reduction of line voltage with dimensional scaling increases processor clock speeds. A possible new low-voltage switching and memory element is the piezoelectric transistor(PET)\cite{NewnsMartyna}. A PET is essentially a transduction device converting the external voltage to stress in a piezoelectric(PE) material which expands\cite{CiureanuMiddelhoek} and in turn compresses a piezoresistive(PR) thin film of the order of a few nanometers\cite{Nalwa} thereby activating a facile insulator-metal transition. Depending on the nature of the transition, continuous or hysteretic, the device can be used as a switch or a memory element, respectively. It is potentially scalable to nanometers and enables device operations at voltages an order of magnitude lower than CMOS while reducing power by two orders and achieving frequencies up to 10 GHz. 

PE materials are well understood and engineered, but the PR materials are more enigmatic. Our work focuses on understanding the electronic origins behind this piezoresistive transition at equilibrium in two primary candidates viz., SmS\cite{SousanisSmet} for PR sensors and memory applications, and SmSe\cite{SousanisPoelman} for switching operations. We also study SmTe belonging to the same family, and look for trends across these materials originating in electronic structure. Experimental observations have indicated that with pressure the bandgap between Sm 4f and 5d bands decreases and an isostructural electronic transition occurs from Sm$^{2+}$ to Sm$^{3+}$ state with an associated mixed valent state discontinuously for SmS\cite{Jayaraman1} and continuously for SmSe, SmTe which outsets a metallic behaviour\cite{Jayaraman2}. SmSe and SmTe show a reversible change in resistivity, but SmS shows hysteresis \cite{MapleWohlleben}$^{,}$ \cite{Deen2005}. 

Samarium(Sm:[Xe]4$^6$6s$^2$) is a lanthanide with open 4f shells. The physical properties of such materials result from an interplay between structure, dimensionality and strong electronic correlations. Sm-monochalcogenides have been studied with first principle approximations to varying degrees over the course of years. While the Density Functional theory (DFT) \cite{HohenbergKohn}$^{,}$ \cite{KohnSham} with local spin-density approximation (LSDA)\cite{Antonov2002} falsely predicts the systems to be metallic while largely underestimating the lattice constants, the LSDA+U approach with a tunable U correctly predicts the bandgaps under ambient conditions but fails to capture the mixed valent high-pressure golden phase of SmS between 6-20 kBar. A self-interaction corrected (SIC) LSDA\cite{Svane2004} shows better predictions for the high pressure intermediate valence state, but needs to correct for the theory-predicted energetics in the trivalent configuration, and describes the system with a total Sm-4f occupation between 5 and 6 by considering it as an array of Sm-f$^5$ ions with one extra partially occupied f-band\cite{Svane2005}$^,$\cite{PetitSvane2016}. Another DFT approach gives better predictions of the lattice constants and bulk moduli by incorrectly making the Sm ions ferromagnetic\cite{Gupta2009}. For the insulating black phase of SmS, DFT+SOC erroneously gives a slightly overlapped semi-metallic state with a valence of 2.56+, and DFT+SOC+U with U=7.6 eV describes correctly a semiconducting state but with a valence of 2.23+\cite{Kang2015}. Many of these studies so far give good bandgap agreements to various degrees but cannot reproduce the multiplet nature of the strongly correlated Sm-4f bands and capture both the states due to the localised nature of partially filled f-orbitals\cite{Ladft} and DFT itself being essentially a one-electron theory.

\begin{figure*}
\includegraphics[width=2.0\columnwidth]{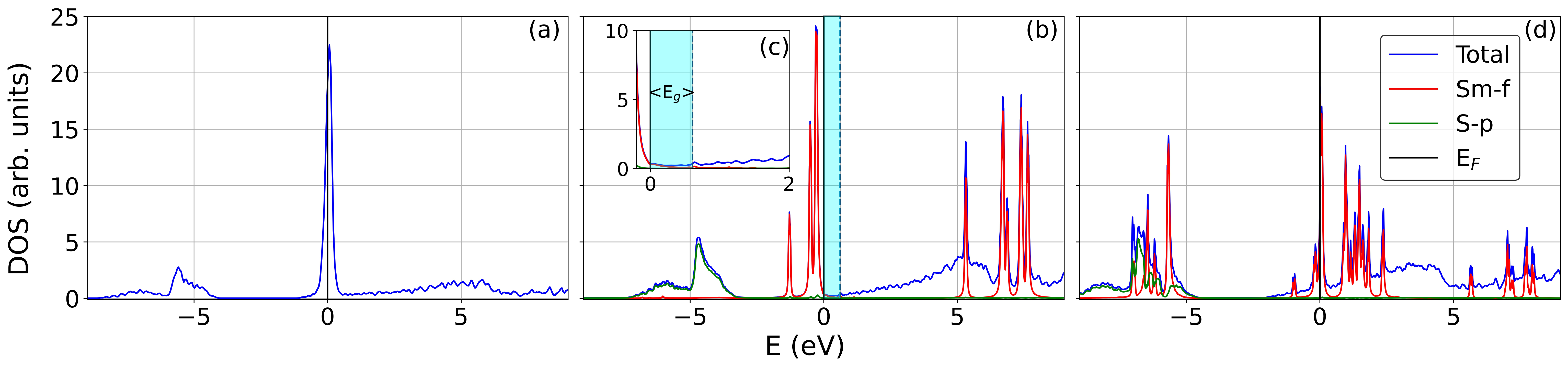}
\caption{Density of states for SmS obtained with (a) DFT, (b) DFT+DMFT for the Sm-4f$^6$ configuration corresponding to ambient pressure insulating black phase, (c) magnified version of (b) showing the existence of a finite bandgap, (d) DFT+DMFT corresponding to the onset of high-pressure metallicity with intermediate Sm-4f valence of 5.24(2) in the golden phase. \label{f1}}
\end{figure*}
The high pressure cohesive properties have also been studied using three-body interaction potential models\cite{MSingh2013}$^,$ \cite{KapoorSingh2017} including long-range Coloumbic forces, short range overlap repulsive forces and polarizability effect\cite{DubeySingh2013} explicitly up to the next nearest neighbour ions. On the other hand, the standard model of lanthanides\cite{LaStandardModel} puts the f-levels in the core. All this still lead to incorrect predictions of conductance and transmission. This entails the necessity of treating the f-levels with a full many-body Hamiltonian beyond DFT, leading to the use of Dynamical Mean Field Theory (DMFT) \cite{MetznerVollhardt}$^{,}$ \cite{GeorgesKotliar}$^{,}$ \cite{GeorgesRozenberg} corrections on top of DFT, and using a DFT+DMFT \cite{KotliarMarianetti} approach  to obtain the correct physics. This realistic application of dynamical mean-field methodology has been widely successful for cases where correlation effects are intrinsically accountable for a phenomenon that cannot be explained by the usual band theory arguments. Phase transitions and crossovers between metallic and paramagnetic insulators driven by temperature as in VO$_2$\cite{goodenough}$^,$\cite{vo2} and V$_2$O$_3$\cite{Rozenberg2}, pressure as for 3d transition metals and compounds\cite{Imada1}, or doping as in Cr-doped V$_2$O$_3$\cite{Limelette89} constitute one class that falls in this category. DMFT has been previously employed with the symmetrized finite-U noncrossing approximation (SUNCA)\cite{Kang2015} to study temperature dependence of SmS and reports a pseudogap appearing near the Fermi level with lowering of temperature after the fashion of Kondo-mixed-valent semimetallic systems. 

\begin{figure*}
\centering
  \includegraphics[width=2.0\columnwidth]{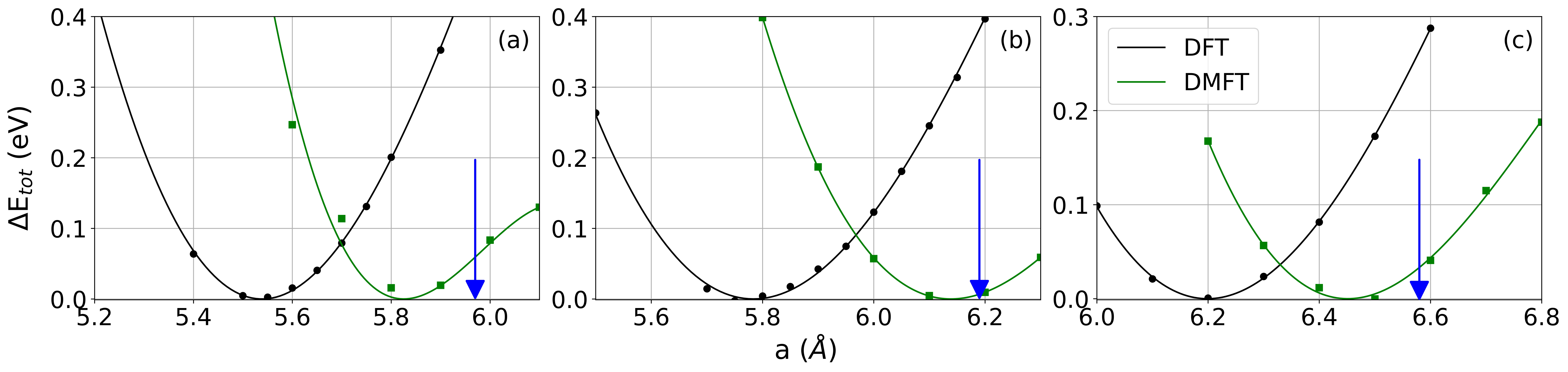}
\caption{Comparison between DFT and DFT+DMFT predictions of total Energy ($E_{tot}$) curves plotted as a function of lattice constant ($a$) and fitted with a Birch-Murnaghan\protect\cite{birch}$^,$\protect\cite{murnaghan} equation of state for (a) SmS, (b) SmSe, and (c) SmTe. Arrows show the experimentally observed values for the respective lattice constants under ambient conditions. \label{f2}}
\end{figure*}

Sm-monochalcogenides(SmX) form fcc-lattices with Sm at (0,0,0) and X(= S, Se, Te) at (0.5,0.5,0.5). The observed isostructural insulator to metal transition in SmX induced by a change in pressure is concomitant with a change in valence which is experimentally characterised by going from Sm$^{2+}$ state at ambient pressure to Sm$^{3+}$ state at higher pressures. Ab-initio calculations using DFT, as implemented in the plane wave basis code CASTEP \cite{RMP-Payne}$^{,}$ \cite{CASTEP}, are done with the PBEsol\cite{pbesol} functional, the FermiDirac smearing scheme with a smearing width of 0.05 eV and a 19$\times$19$\times$19 Monkhorst k-grid\cite{MPgrid}. They always converge to a metallic ground state with Sm-4f peaks at the fermi level (Fig.\ref{f1}(a)). Consistent with the literature, putting a tunable U$_{{eff}}$ value shifts the Sm-4f levels accordingly and converges to the correct insulating ground state. It however fails to reproduce the metallic state at higher pressures. As there is a change in valence, it is likely that this metal-insulator transition (MIT) is accompanied by intermediate valence states which are not well captured by DFT, it being in principle a single Slater determinant approach. The presence of strongly correlated Sm-4f peaks at the fermi level are key to these experimentally observed intermediate valence states leading to the onset of metallicity. To capture this fractional occupation state, we look beyond such mean-field approximations warranting the need for many body corrections.

DMFT has traditionally been a good tool to describe such MITs. Since the 4f-series is usually characterized by a small hybridization, DMFT is used here within the scope of the Hubbard-I(HI) approximation. Thus, the Hubbard-I solver\cite{h1} as implemented in CASTEP is used with the ensemble density functional method to perform DMFT calculations at various values of lattice constant and valence. We set the inverse temperature $\beta = 20$, and use 2048 Matsubara frequencies to calculate the Green's functions in the fully localized limit(FLL) approximation for the double counting correction scheme. We use a standard value of U=6.1 eV on Sm and a Hund's coupling of 0.3 eV. We find that for the correct ordering of X-p levels with respect to the Sm-4f states in the metallic phase, a static DFT-U of 6 eV has to be put on X. For the insulating state this U does not change the physics qualitatively around the Fermi level other than an expected small shift in the occupied p-states, and thus has been used throughout consistently.

DMFT gives us a control over the degree of correlation and valence of the system across this transition. We capture the effect of pressure via the nominal occupancy($n_{if}$) which characterises the Sm-4f electronic valence and is controlled by the double counting correction as implemented in the HI solver. We then perform energy minimisations to reach the electronic ground state. A clear distinction between insulating (Fig.\ref{f1}(b, c)) and metallic (Fig.\ref{f1}(d)) behaviours emerge for SmS across the MIT as the system goes from ambient to high pressure. This is accompanied by a charge difference of 0.76 which confirms the presence of an intermediate valence state for this material and matches reasonably well with the experimentally observed values of 0.62\cite{Deen2005}$^,$\cite{handbookval}. The density of states (DOS) for the SmS black phase matches well the observed ARPES spectra\cite{2002arpes}. Additionally with DMFT corrections, the subtle effects due to the paramagnetic nature of SmS also emerge associated with the presence of a local fluctuating magnetic moment between 1.96-2.12 $\mu$B, not well captured by DFT.

\begin{table}
\begin{center}
\vline
\begin{tabular}{c|c|c|c|c|c|c}
\hline
Material & $a_{exp}$ & $a_{DFT} $ & $a_{DMFT} $ & $n_{f}$ & $B_{0-exp}$ & $B_{0-DMFT}$  \\
 & (\AA) & (\AA) & (\AA) & & (GPa) & (GPa) \\
\hline
\hline
SmS &  \cellcolor{red}5.97 & \cellcolor{green}- & \cellcolor{red}5.82 & \cellcolor{red}6.00(2) & 89.8-92 & 90.6 \\
\hline
 & \cellcolor{green}5.7 & \cellcolor{green}5.54 & \cellcolor{green}5.68 & \cellcolor{green}5.24(2) & - & - \\
\hline
SmSe & \cellcolor{red}6.19 & \cellcolor{green}5.78 & \cellcolor{red}6.14 & \cellcolor{red}6.00(2) & 40$\pm$5 & 62.5 \\
\hline
SmTe & \cellcolor{red}6.58 & \cellcolor{green}6.2 & \cellcolor{red}6.45 & \cellcolor{red}6.00(2) & 40$\pm$5 & 49.5 \\
\hline
\end{tabular}\vline
\caption{ Comparison of experimental data\protect\cite{SousanisSmet}$^,$\protect\cite{handbookval}$^,$\protect\cite{handbook}$^,$\protect\cite{1995bihan} with DFT and DMFT predictions for lattice constant ($a$) and bulk-modulus ($B_0$) for SmS, SmSe, and SmTe, along with DMFT predicted Sm-4f electron occupation ($n_{f}$). The red and the green cells of $a$ and $n_{f}$ stand for insulating and metallic states respectively, either predicted or observed. \label{t1}}
\end{center}\vspace{-8mm}
\end{table}
\begin{figure*}
\centering
\includegraphics[width=2.0\columnwidth]{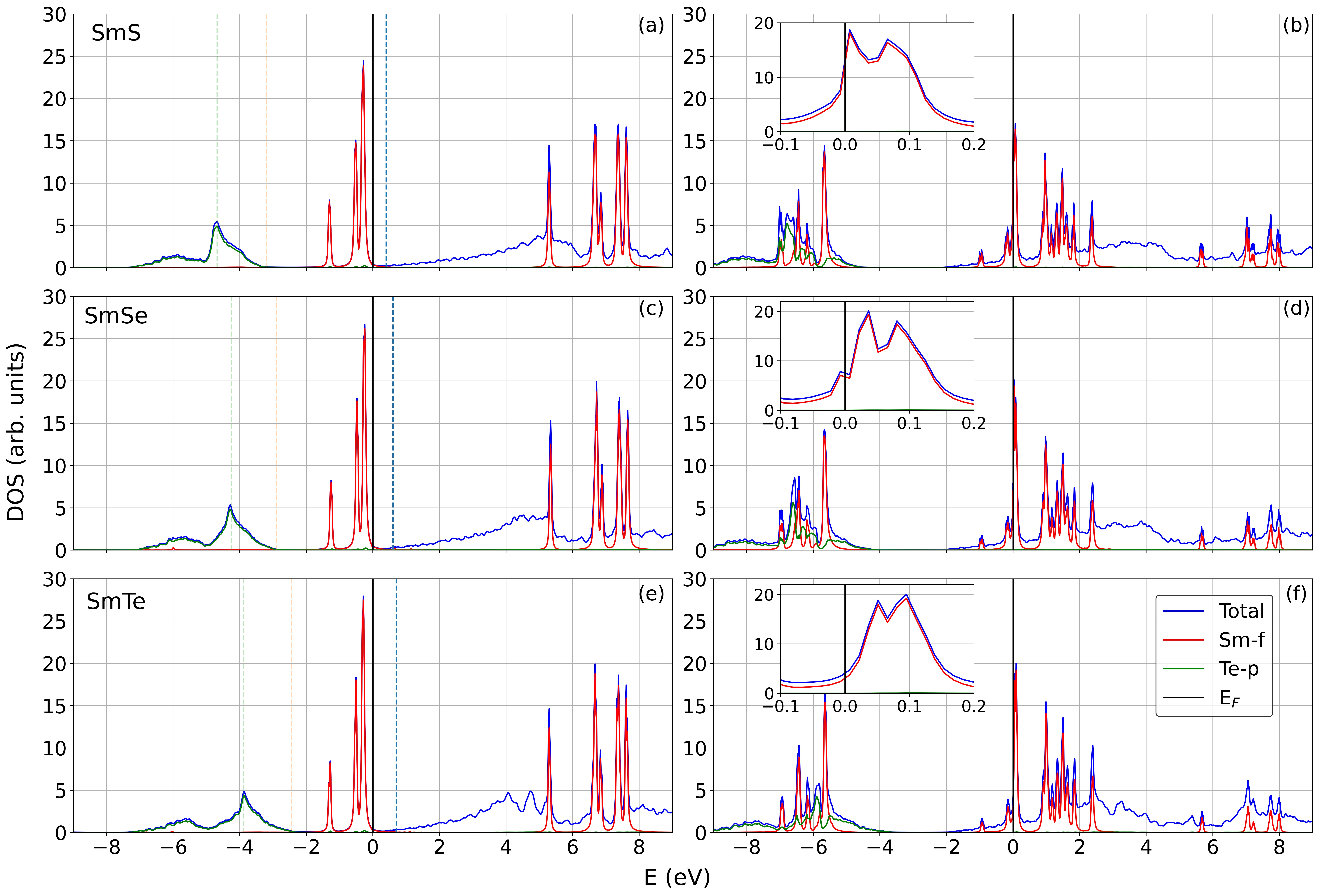}
\caption{(\textit{Left column}) Density of states obtained with DFT+DMFT corresponding to the ambient pressure insulating states for (a) SmS, (c) SmSe, and (e) SmTe, and (\textit{right column}) the high-pressure intermediate valence metallic states with insets focusing on the behaviour of Sm-4f peaks around the fermi level for (b) SmS, (d) SmSe, and (f) SmTe, respectively. \label{f3}}
\end{figure*}
\begin{figure*}
\centering
  \includegraphics[width=2.0\columnwidth]{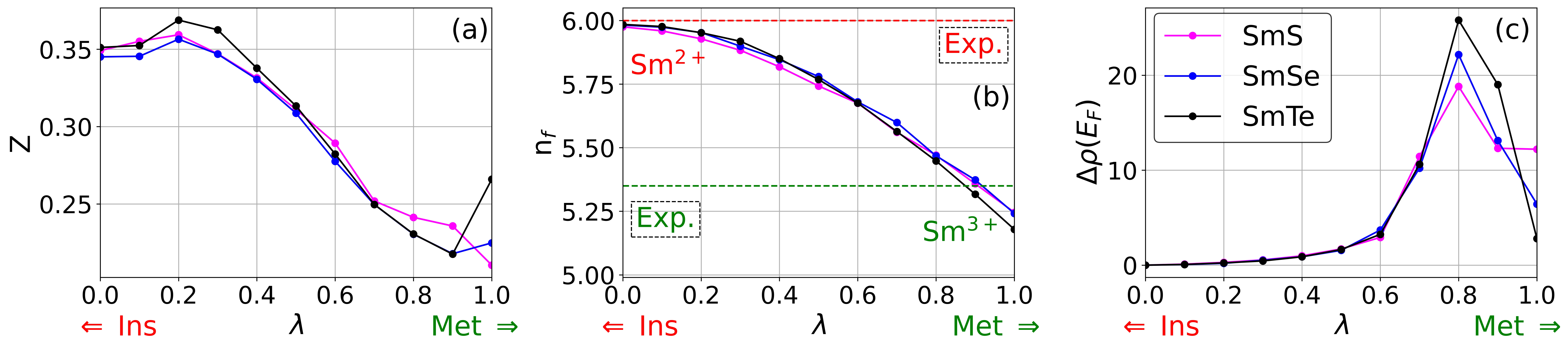}
\caption{(a) Total quasiparticle weight ($Z$), (b) Sm-$4f$ electron-occupancy ($n_{f}$), (c) total density of states at the Fermi level ($\rho(E_F)$) along the NEB path going from the insulating state ($\lambda=0$) to the metallic state ($\lambda=1$) for SmS, SmSe and SmTe. \label{f4}}
\end{figure*}
We minimise DMFT obtained total energies to relax and optimise the structures which provide the lattice parameters and Birch-Murnaghan\cite{birch}$^,$\cite{murnaghan} fitted values of bulk moduli reported in Table \ref{t1}, and show significant quantitative improvements over DFT predictions, and agree well with experiments\cite{SousanisSmet}$^,$\cite{1995bihan} (Fig.\ref{f2}) for all three SmX materials. At the onset of the golden phase of SmS we predict a non-integer occupancy of 5.24(2), and of 6.00(2) at the beginning of all the insulating states. Note however that in contrast with previous studies where long range magnetic order\cite{Gupta2009}, or spin-orbit coupling\cite{Kang2015} or next-nearest neighbour interactions\cite{DubeySingh2013}$^{,}$\cite{KapoorSingh2017} were invoked to reproduce experiments, in our work we recover consistent results with experimental observations without the need of further mechanisms at room temperature. Although PBEsol underestimates the lattice parameters, we obtain remarkable experimental agreements arising from local charge and spin fluctuation corrections using DMFT.

In the insulating states we find the bandgap to increase by approximately 50\% and then 20\% when going down from isovalent SmS to SmTe (Fig.\ref{f3}(a, c, e)). However even with increasing mass and more electrons, the qualitative natures of the DOS remains essentially the same with some little changes mostly regarding the occupied X-p levels shifting towards the fermi level. For the metallic states, we observe the sharp 4f-peaks to be almost on top of the fermi level within 0.01 eV for SmS, and just across it within 0.1 eV for SmSe and SmTe shifting towards the right as we go down the group (Fig.\ref{f3}(b, d, f)). These shifts can be attributed to the previously noted increase in bandgap which pushes the f-states further above the fermi level, and is consistent with experimental observations\cite{1995bihan}, and might explain the difference between the discontinuous nature of the MIT in SmS and their continuous nature in SmSe, SmTe. This very sharp resonance is crucial to the onset of metallicity as small differences in pressure, temperature, disorder might shift this peak to left or right and cause huge changes in the density at the fermi level making its position a critical parameter for transport and calorimetry. A point to note here is that the metallic states for SmSe and SmTe have not been observed in experiments. This can be regarded as a utility of DMFT as we can hereby predict properties of systems which are not fully stabilised in experiments, and can use this to gauge the effect of mass contrast in these systems.

To investigate the nature of the MIT further, we employ the nudged elastic band (NEB) method\cite{neb}. We select a path from the predicted DMFT solution for the insulating states corresponding to the larger value of lattice constants, $a$, under ambient pressure with the NEB-parameter $\lambda=0$ and interpolate it to what we predict to be the high-pressure metallic states corresponding to $\lambda=1$. Along this path, while going from the insulating to the metallic phases, there is a regime where the f-states cross the fermi level (Fig.\ref{f4}(c)). This variation of density corresponds entirely to the impurity states confirming that the metallic conductivity in the SmS golden phase originates from the Sm-4f states as seen in experiments \cite{2013arpes}$^,$\cite{1975hall}. Additionally along this path, we observe a change in occupancy which follows well the experimentally predicted valence for the insulating and the intermediate valence metallic states (Fig.\ref{f4}(b)). This trend in valence remains similar for all three materials. Note that as our NEB approach interpolates linearly between the structural and electronic properties of the respective metallic and insulating phases, we cannot infer about possible first order transitions or sharp discontinuity in the Sm valence.

But contrary to expectations, not only the quasiparticle weight($Z$) decreases, but is also non-zero for the observed insulating phases (Fig. \ref{f4}(a)). However, when we reach the predicted metallic states for SmSe and SmTe, $Z$ goes up a little. Although the effective mass is mostly dominated by the mass renormalisation due to the f-electrons, it shows that this MIT is not a Mott transition. Based on this behaviour, we argue that these are correlated band insulators instead, confirming ARPES results for SmS\cite{2013arpes}, where on driving towards the metallic phase the Sm-4f interestingly become even more correlated, and thus we conclude that here $Z$ is not a measure of transition. In the predicted metallic states, the f-states are weakly hybridised at the fermi level. Across the MIT, they do cross the fermi level but do not end up in the metastable configuration. So, the 4f-states are across the energy barrier that separates these two phases. At some point along the NEB, we are passing through a state which is reminiscent of Kondo lattices. These f-states are however very important for transport and the role of SmX as PR materials because if we gate the system, small voltages might suddenly involve the f-states in the conduction and they might start contributing more.

In conclusion, the inability of DFT to explain the pressure induced insulator to metal transition with a concomitant change in valence in Sm-monochalcogenides is corrected using DMFT to access fractional electronic occupations. It leads to significant improvements over the predicted values of lattice constants and bulk-moduli for the experimentally observed states as a direct result of incorporating the charge and spin fluctuations. We also obtain the multiplet nature of the DOS with excellent experimental agreements in absence of any long range interactions or spin-orbit coupling. On doing NEB, we find that the quasiparticle weight remains finite for the insulating phases and decreases with pressure, and is not an indication of this transition, thus inferring that this is not a Mott transition. As a result we classify these materials as correlated band insulators.

\section*{Acknowledgements}

DB acknowledges the support of the NPIF grant [EP/R512552/1] and the UK EPSRC CDT for CANES [EP/L015854/1]. CW and EP were supported by the grant [EP/R02992X/1] from EPSRC. IR acknowledges the support of the UK department for BEIS through the UKNQT programme. DB and IR were also supported by the EU Horizon2020 research and innovation programme within the PETMEM project [No.688282].

This work was performed using resources provided by the ARCHER UK National Supercomputing Service and the CSD3 facility hosted by the University of Cambridge [Capital grant EP/P020259/1].

\section*{References}
\nocite{*}
\bibliographystyle{apsrev}
\bibliography{SmXPRL}

\end{document}